# Energy- Angular Correlation of Medium Energy Particles Produced in Heavy Ion Collisions

M.T. Hussein, N.M. Hassan, N.M. Sadek and Jamila Elsweedy
Physics Department, Faculty of Science, Cairo University, 12613 Giza, Egypt

Abstract

The nuclear photo-emulsion technique is used to study the information carried by the medium energy nucleons produced in heavy ion collisions. Multiplicity, energies as well as the angular distribution of this type of particles are measured. Due to the difficulties in measuring the energy only some particles having special criteria could be selected to measure their energy with consenting accuracy. A hypothetical model is proposed to correlate the energy of the produced particles to their emission angles so that it becomes easy to estimate the energy distribution in terms of measured emission angle. The proposed model is constructed upon statistical thermodynamic assumptions. Moreover, two additional base functions are originated that play the role of the statistical angular weight factor and the nuclear density of the compressed nuclear matter at the moment of particle emission. The prediction of the model are compared with complete set of measured data of the reactions of proton, helium, carbon and neon nuclei with the composite emulsion nuclei as target at 4.2 A GeV.

## 1. Introduction

The nuclear emulsion is a good tool in dealing with high energy nuclear reaction. It has the ability to detect and identify particles in the outlet channel of the reaction. The dynamic characteristics of the reaction can be determined perfectly by resume precise measurement of the angular distribution, the energy spectra as well as the charge distribution of the produced particles which carry information about the mechanism of the interaction. Unfortunately, the measurements of the energy of charged particles is a tedious work and requires pursuing their path through the emulsion plates for enough long distance to get accurate results. On the other hand it is possible to measure the angular distribution to a high extent of accuracy. In this intellect we believe that finding a kind of correlation between the angular and energy distribution of the emitted particles will be good evidence in improving the performance of using the emulsion as a tool in determination of the energy spectra of the produced particles. In this work we aim to weave a scenario that discuss what happen inside the strong interacting nuclear matter and get the link between the energy and angular behavior of the emitted particles in the terminology of thermodynamics. The predictions of the model will be compared with the experimental data of the reactions of proton, Helium, Carbon and Neon nuclei of the emulsion target at 4.2 A GeV incident energy where the energy of the gray particles are well measured.

## 2. The Formulation of the Model

It is assumed that during the collision of fast projectile nucleus **P** with a target **T** at a given impact parameter b, a large amount of energy is transferred from the projectile to the target and the nucleons of both nuclei defuse through each other. It is plausible to work with a parameter that defines the fraction of the projectile nucleons in the formed nuclear system as:

$$\eta(b) = \frac{\rho_p(b)}{\rho_p(b) + \rho_T(b)} \qquad (1)$$

where $\rho_p(b)$ and $\rho_T(b)$ are the projectile and target densities at a given impact parameter $b$ in the formed nuclear matter. $\eta$ has a continues values extending from zero to 1. It is zero in the target region and goes to 1 as we approach the projectile region. It is possible to imagine three separate regions without clear boarders as shown in Fig.(1). These are the projectile spectator, the target spectator and an overlap region. The parameter $\eta$ plays an important role in understanding the physics inside each part of the interacting medium. The quantity of energy transferred and the activity of nuclear collisions whether it was a strong collision or even elastic or Coulomb dissociation is controlled by the value of $\eta$.

**The projectile spectator region:** is characterized by small momentum transferred that is enough to dissociate the projectile into few fragments moving in the forward direction or scattered by relatively small angle. Simple elastic scattering [1] assuming ptical potential [2], diffraction [3] and Coulomb dissociation [4] models are sufficient to describe the fragmentation process and the angular spread of the emitted fragments in this region.

**The target spectator region:** The nucleons in this region are initially at rest. As the collision starts up, nucleons from the projectile defuse slowly through the target transferring a little bit fraction of the projectile energy. The diffusion rate depends mainly on the impact parameter. The system then behaves as perfect gas that suffers multiple of successive elastic scattering. Consequently the entropy of the system increases until it reaches equilibrium state, with equilibrium temperature of the order of 30 MeV. At this moment the system evaporates [5] producing heavily ionizing fragments appear as black particles with isotropic distribution in the space. In most cases it was sufficient to describe the energy distribution of the evaporated particles with a unique Maxwell distribution of classical distinguished particles.

**The hot spot region:** The overlap region between the projectile and the target that characterizes with $\eta \approx 0.5$ is the hottest region in the space. Large amount of heat is dissipated there. The nuclear matter goes through different stages. In the early one a sudden compression occurs to the nuclear matter accompanied by much increase in matter density and production of large amount of center of mass energy. The environment is now adequate for the formation of quark-gluon plasma phase [6]. Many quarks-antiquarks are being created followed by a recombination process. The created quark pairs form what are called sea quarks. Neighboring quark-antiquark may recombine again forming meson [7]. Successive collisions go on producing more newly created particles and hence the system expands again until the collisions cease. If we treat the system thermodynamically [8], it is expected that fast light particles be produced in the early stage in the forward direction. As time goes up, the system is subjected to successive collisions each of them followed by creation of bunches of newly produced particles that emitted in wider emission angle. Finally the nucleons are emitted individually or rather in cluster or fragment form. The singly charged fragments are produced as knocked on nucleons with medium energy range ($40\,MeV < E < 400\,MeV$) and appear in emulsion plates as gray particles [9]. In the present work we are interesting with this type of particles. We treat the nuclear matter as a nonequilibrium system. Each point in the space is considered as local equilibrium subsystem behaves as a canonical ensemble that is characterized by a specific temperature and a specific projectile fraction parameter $\eta$. The overall distribution of the gray particles is found by the superposition of particles produced over the assembly of the different subsystems covering the range of $\eta$. It is also assumed that particles are produced isotropically in the center of mass of each ensemble showing Maxwell Boltzmann distribution for classical particles, Fermi-Dirac for fermions and Bose-Einstein for bosons [10]. On the other hand, since the center of mass itself is moving with a velocity with respect to the Lab system related to its $\eta$ value then the emitted particles are produced with anisotropic decay. The degree of anisotropy depends on the center of mass velocity or the energy of the emitted particles. Our goal is to get a correlation relation between the energy and the angular spread of the emitted particles. We use Gaussian density distribution for nuclei with A<20 and Woods-Saxon for A>20 [11]. Using appropriate units where $\hbar = c = k = 1$, then the center of mass energy dissipated in a local position is given by:

$$\zeta_{cm} = 3T + m\frac{K_1(m/T)}{K_2(m/T)} \quad (2)$$

The conservation of energy at a given location requires that:

$$[m^2 + 2\eta(1-\eta)mt_i]^{1/2} = 3T + m\frac{K_1(m/T)}{K_2(m/T)} \quad (3)$$

m is the nucleon rest mass, $t_i$ is the incident kinetic energy per nucleon in the Lab system and $K_1$ and $K_2$ are the Mc Donald's functions of first and second orders respectively. The solution of Eq.(3) results the value of the local temperature at the specific η value. The variation of the temperature with η is displayed in Fig.(2). Maximum temperature corresponds to $\eta = o.5$. The temperature devolves towards both the projectile and the target regions. It is assumed that at each local equilibrium point the gray particles are produced in Maxwellian form subjected to the corresponding temperature.

$$F(E,\eta) = \frac{d^2N}{p^2 dp d\Omega}$$
$$= \frac{N}{4\pi m^3} \frac{Exp(-E/T)}{2(T/m)^2 K_1(m/T) + (T/m)K_0(m/T)} \quad (4)$$

Eq.(4) describes the energy distribution of the gray particles produced in the rest frame of the hot spot nuclear matter which shows isotropic distribution there. Transforming this distribution to the Lab system, assuming that the nuclear source is moving with velocity $\beta_{cm}$ with respect to the Lab system, hence the produced gray particles are emitted with angle $\theta_L$ in the Lab system.

$$E = \gamma_{cm}(E_L - \beta_{cm} P_L \cos\theta_L) \quad (5)$$

$$\beta_{cm} = \frac{P_L}{E_L} = \frac{\eta[(t_i + 2m)]^{1/2}}{m + \eta t_i} \quad (6)$$

The Lab distribution function $F_L(E_L,\eta,\theta_L)$ describes the energy distribution of the emitted gray particles from a source with certain η at a given Lab angle $\theta_L$. The energy distribution in the lab system is found by integration over η and $\theta_L$ each of them weighted by the corresponding statistical weight factor. The η weight factor depends mainly on the density distribution of the interacting nuclei, their diffuseness and their temporary compressibility at the moment of emission. The detailed formulation of this factor is much complicated. The global effects of these factors are considered in an exponential parametric form $\chi(\eta)=Exp(-\delta\eta)$. The parameter δ carries information about the geometry of the system and its compressibility and the diffuseness shape of the matter density of the interacting nuclei. Data Manipulation package [12] is loaded from the *Mathematica* software to find the best values of δ that fit the experimental data. Table (1) shows that the value of the parameter δ decreases rapidly with the projectile mass. The target mass is fixed and considered as the average value of the composite emulsion nuclei. On the other hand, the angular weight factor is to be lending from the experimental results since the angular distribution $Y(\theta)$ is measured in emulsion technique to a high extent of accuracy with sufficient confidence. In Fig.(3) we display the angular distribution of gray particles produced at the same incident energy 4.2A GeV, for the projectiles proton, helium, carbon and neon interaction with emulsion nuclei. The result shows that almost all the distributions come close to each other which support the idea that the target is the source of the gray particles. Fig.(4) shows the energy distribution of the gray particles produced at fixed angle $\theta=\pi/6$ at low η values (Fig. (4-a)) and high η values (Fig. (4-b)). At low η values the temperature is enough small so that most of the distribution area is covered within the range of the gray particles (30 – 400 MeV), while the curves drawn at high η describe only the front portion of the Maxwell distribution just before recognizing the peak of the curve and the left portion corresponds to fast particles with energy more than 400 MeV, those appears in emulsion as shower particles.

The final form of the energy distribution of the gray particles is found by integration over the η and all the θ range so that:

$$F_L(t) = \int_0^{2\pi}\int_0^1 F(t,\eta,\theta)\chi(\eta)Y(\theta)d\eta d\theta \qquad (7)$$

The energy $E$ in Eq. (7) is replaced by the corresponding kinetic energy t, $E=m+t$, to put the relation in an appropriate form for comparison with the available experimental data. The predictions of Eq.(7) for the reactions of p, He, C and Ne at 4.2 AGeV with emulsion target are shown in Fig.(5) compared with experimental data [13,14] where fair agreement is obtained. This result give us an authorized certitude to apply Eq. (7) successfully to get the energy distribution of gray particles produced within an angular band just we know the angular distribution $Y(\theta)$ and the compressed density factor $\chi(\eta)$.

| Reaction | δ |
|----------|-----|
| p-Em | 0.9 |
| He-Em | 0.3 |
| C-Em | 0.2 |
| Ne-Em | 0.1 |

TABLE (1) The diffuseness parameter δ of the compressed nuclear matter in the hot spot region as predicted by the model.

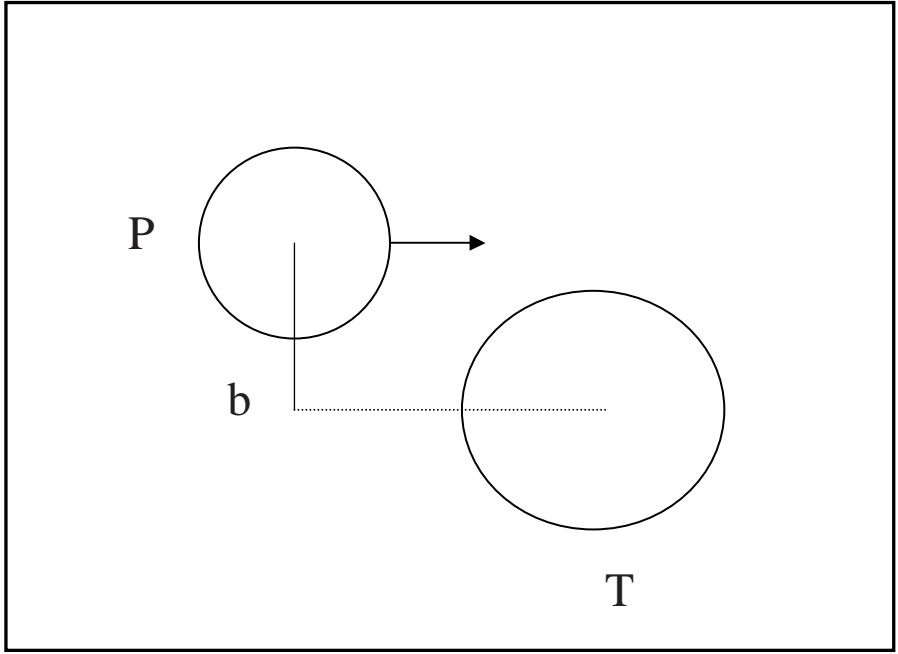

Fig (1-a)

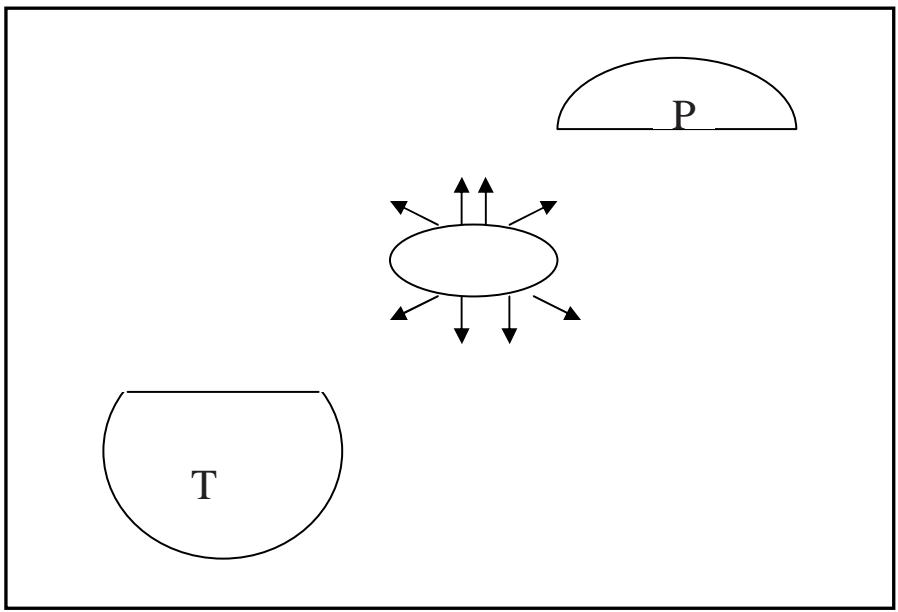

Fig (1-b)

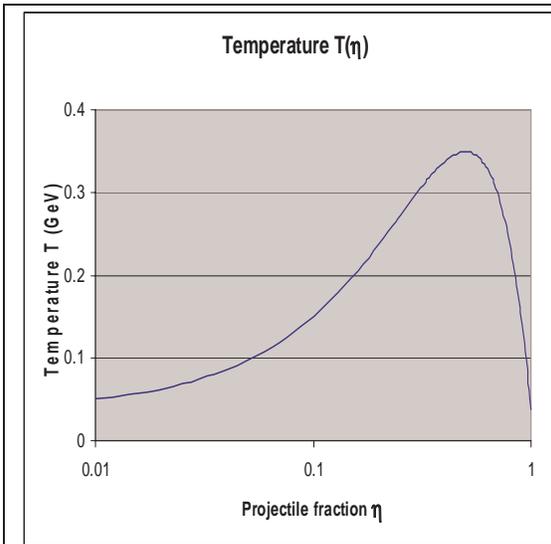

Fig.(2) The temperature T of the thermo-dynamic system as a function of the projectile fraction η.

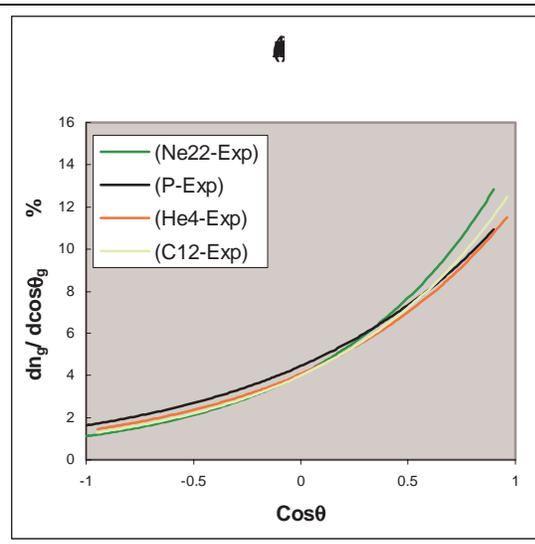

Fig.(3) The angular distribution of grey particles produced in (P-He-C-Ne) interactions with Emulsion nuclei at momentum 4.5 GeV/c for P-He-C per nucleon and 4.1 GeV/c for Ne per nucleon.

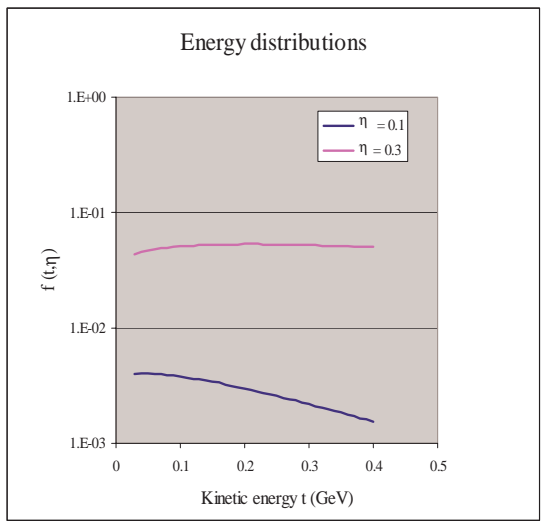

(a)

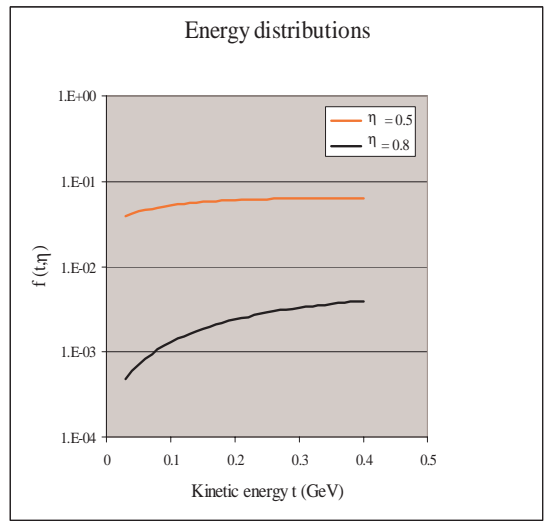

(b)

Fig.(4) Energy distribution of medium energy protons produced at angle θ=π/6 at projectile fraction (a) η= 0.1, 0.3 and (b) η=0.5, 0.8

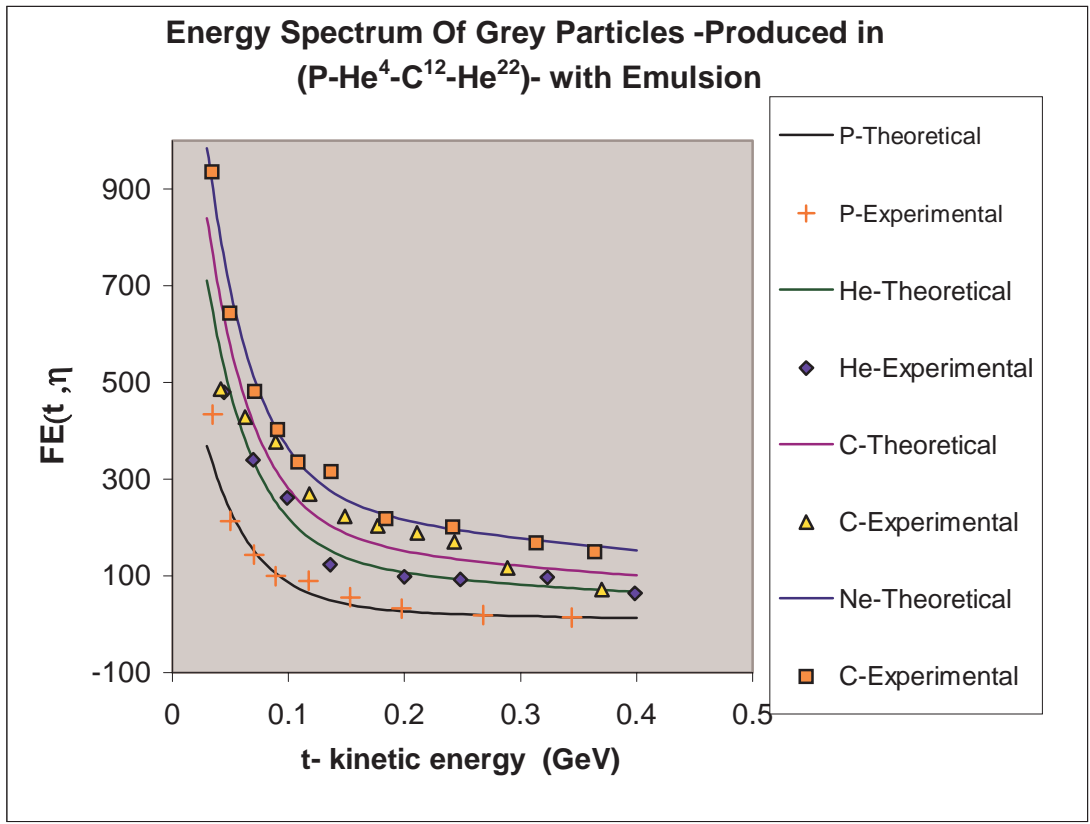

Fig. (5) The energy distribution of gray particles produced in p-Em, He-Em, C-Em and Ne- Em at 4.2 A GeV. The solid line is the prediction of the thermodynamic model which is calculated in terms of the measured angular distribution of the corresponding reaction. The nuclear matter density factor is taken as in Table (1).